\documentclass[aps,pra,twocolumn,superscriptaddress,nobalancelastpage]{revtex4-1}

\usepackage{amsmath,amssymb,mathtools}
\usepackage[dvipdfmx]{graphicx}
\usepackage[colorlinks=true,bookmarks=true,citecolor=blue,linkcolor=blue,urlcolor=blue,breaklinks=true]{hyperref}

\graphicspath{{fig/}}
\usepackage{dcolumn}
\usepackage{bm}
\usepackage[T1]{fontenc}
\usepackage{textcomp}
\usepackage{color}
\usepackage{mathrsfs}
\usepackage[normalem]{ulem}
\usepackage{enumerate}
\usepackage{setspace}

\begin{document}

\title{Non-Hermitian Casimir effect of magnons}

\author{Kouki Nakata}
\email[(Equal contribution) \ ]{{nakata.koki@jaea.go.jp}}
\affiliation{Advanced Science Research Center, Japan Atomic Energy Agency, Tokai, Ibaraki 319-1195, Japan}

\author{Kei Suzuki}
\email[(Equal contribution) \ ]{{k.suzuki.2010@th.phys.titech.ac.jp}}
\affiliation{Advanced Science Research Center, Japan Atomic Energy Agency, Tokai, Ibaraki 319-1195, Japan}

\date{\today}

\begin{abstract}
\noindent{\textbf{Abstract.}}
There has been a growing interest in non-Hermitian quantum mechanics. The key concepts of quantum mechanics are quantum fluctuations. Quantum fluctuations of quantum fields confined in a finite-size system induce the zero-point energy shift. This quantum phenomenon, the Casimir effect, is one of the most striking phenomena of quantum mechanics in the sense that there are no classical analogs and has been attracting much attention beyond the hierarchy of energy scales, ranging from elementary particle physics to condensed matter physics, together with photonics. However, the non-Hermitian extension of the Casimir effect and the application to spintronics have not yet been investigated enough, although exploring energy sources and developing energy-efficient nanodevices  are its central issues. Here we fill this gap. By developing a magnonic analog of the Casimir effect  into non-Hermitian systems, we show that this non-Hermitian Casimir effect of magnons is enhanced as the Gilbert damping constant (i.e., the energy dissipation rate) increases. When the damping constant exceeds a critical value, the non-Hermitian Casimir effect of magnons exhibits an oscillating behavior, including a beating one, as a function of the film thickness and is characterized by the exceptional point. Our result suggests that energy dissipation serves as a key ingredient of Casimir engineering.
\end{abstract}

\maketitle

\section{Introduction}
\label{sec:intro}

Recently, 
non-Hermitian quantum mechanics
has been drawing considerable attention~\cite{Review_NH_Ashida}.
The important concepts of quantum mechanics are 
quantum fluctuations. 
Quantum fluctuations of quantum fields 
under spatial boundary conditions
realize a zero-point energy shift.
This quantum effect which arises from the zero-point energy,
the Casimir effect~\cite{CasimirEffect,CasimirExp,CasimirExpErra,CasimirExpPRL2002},
is one of the most striking phenomena 
of quantum mechanics in the sense that 
there are no classical analogs.
Although the original platform for the Casimir effect~\cite{CasimirEffect,CasimirExp,CasimirExpErra,CasimirExpPRL2002} 
is the photon field~\footnote{See
Ref.~\cite{CasimirOscillation}, as an example,
for an oscillating behavior of the Casimir effect
of photons
as a function of distance
between two uncharged plates,
where chiral material
inserted between the two parallel plates plays a key role.},
the concept can be extended to various fields 
such as scalar, tensor, and spinor fields~\cite{CasimirCosmoloReview,CasimirReview1986,CasimirReview1988,CasimirReview2001,CasimirNanoRMP,ReviewCasimirNatPhoto,Casimir_YM_Maxim_PRL2018,Casimir_YM_Maxim_PRD2019,Casimir_YM_Kitazawa_PRD2019}.
Thanks to this universal property, 
the Casimir effects have been investigated
in various research areas~\footnote{As an example, see Refs.~\cite{AFCasimir1989,CasimirAF1992,AFmagnonCasimir1998,SpinCasimir2015,CasimirFM,SkyrmionFMCasimir,IvanovCasimirSoliton}
for Casimir effects in magnets
and Ref.~\cite{RChengAFthermalCasimirMagnon}
for a magnonic analog of the thermal Casimir effect
in a Hermitian system.
For details of the distinction 
between  
the thermal Casimir effect
and
the Casimir effect,
refer to Supplemental Material.
See also Ref.~\cite{CasimirDynamicalSaito} for an analog of the dynamical Casimir effect with magnon excitations in a spinor Bose-Einstein condensate.}
beyond the hierarchy of energy scales~\cite{CasimirCosmoloReview,CasimirReview1986,CasimirReview1988,CasimirReview2001,CasimirNanoRMP,ReviewCasimirNatPhoto,Casimir_YM_Maxim_PRL2018,Casimir_YM_Maxim_PRD2019,Casimir_YM_Kitazawa_PRD2019}, 
ranging from elementary particle physics 
to condensed matter physics,
together with photonics.
However, 
the non-Hermitian extension of the Casimir effect
and the application to spintronics
remain missing ingredients,
although exploring energy sources
and developing the potential for energy-efficient nanodevices 
are the central issues of spintronics~\cite{LLGspintroReview,MagnonSpintronics,Review_QuantumMagnonics,NHreviewBenedetta,ReviewNHmagnonics}.

\begin{figure}[t]
\centering
\includegraphics[width=0.515 \textwidth]{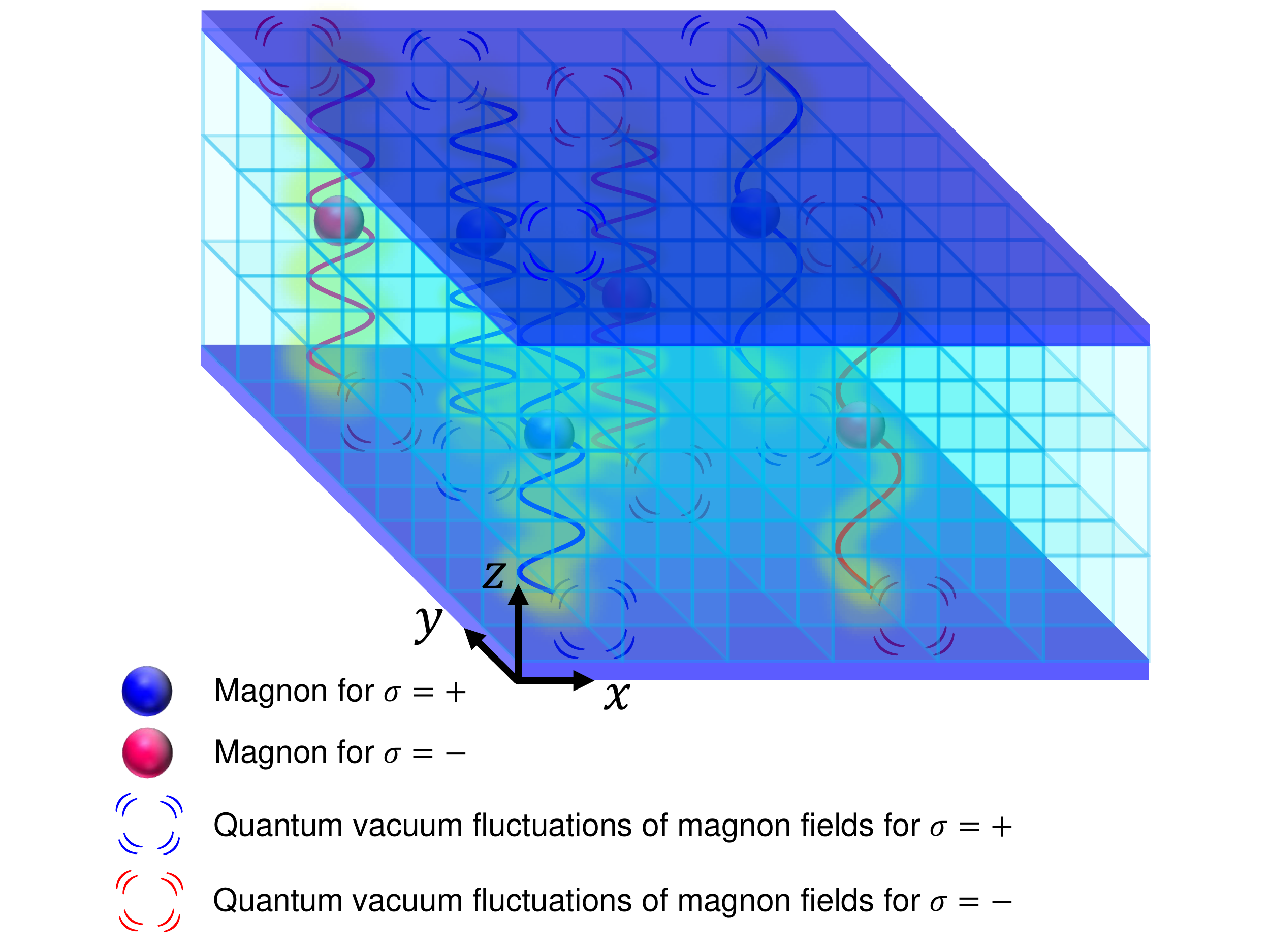}
\caption{{Schematic of magnonic Casimir effect.
Magnonic Casimir effect arises 
from quantum vacuum fluctuations of magnon fields.}
}
\label{fig:system}
\end{figure}

Here we fill this gap.
The Casimir effects are characterized by
the energy dispersion relation.
We therefore incorporate
the effect of energy dissipation on spins
into the energy dispersion relation of magnons
through 
the Gilbert damping constant~\cite{GilbertOriginal}
and thus 
develop a magnonic analog of the Casimir effect~\cite{magnonCasimir_KK}, 
called the magnonic Casimir effect
(see Fig.~\ref{fig:system})~\footnote{In Ref.~\cite{magnonCasimir_KK},
we investigated the Casimir effect 
induced by quantum fields for magnons
(i.e., a magnonic analog of the Casimir effect) 
and referred to it as the magnonic Casimir effect.
See Ref.~\cite{magnonCasimir_KK} 
for details of the magnonic Casimir effect 
in dissipationless systems.},
into non-Hermitian systems.
We then show that 
this non-Hermitian extension of the magnonic Casimir effect,
which we call 
the magnonic non-Hermitian Casimir effect,
is enhanced as the Gilbert damping constant increases.
When the damping constant exceeds a critical value, 
the magnonic non-Hermitian Casimir effect
exhibits an oscillating behavior
as a function of the film thickness
and is characterized by 
the exceptional point~\cite{EP_kato} (EP).
We refer to this behavior as
the magnonic EP-induced Casimir oscillation.
We emphasize that 
this magnonic EP-induced Casimir oscillation
is absent in the dissipationless system of magnons.
The magnonic EP-induced Casimir oscillation
exhibits a beating behavior in the antiferromagnets (AFMs)  
where the degeneracy between two kinds of magnons is lifted.
Our result suggests that energy dissipation serves 
as a new handle on
Casimir engineering~\cite{Review_CasimirEngineering}
to control and manipulate 
the Casimir effect of magnons.
Thus, we pave a way for magnonic Casimir engineering
through the utilization of energy dissipation.

\section{Results}
\label{sec:II}

\subsection{System}
\label{subsec:II}

We consider 
the insulating AFMs of two-sublattice systems
in three dimensions
described by 
the Hamiltonian,
\begin{equation}
{\mathcal{H}}=
J \sum_{\langle i,j \rangle}
{\mathbf{S}}_i \cdot {\mathbf{S}}_j
-K_{\text{e}}  \sum_{i} (S_i^y)^2
+K_{\text{h}}  \sum_{i} (S_i^x)^2,
\label{eqn:H}
\end{equation}
where
$ {\mathbf{S}}_i=(S_i^x, S_i^y, S_i^z)  $
is the spin operator
at the site $i$,
$J > 0$ parametrizes the antiferromagnetic exchange interaction 
between the nearest-neighbor spins
$ \langle i,j \rangle$,
$K_{\text{h}}>0$ is the hard-axis anisotropy,
and $K_{\text{e}}>0$ is the easy-axis anisotropy.
These are generally
$K_{\text{h}}/J \ll 1$
and
$K_{\text{e}}/J \ll 1$.
The AFMs have the N\'eel magnetic order
and there exists
the zero-point energy~\cite{AndersonAF,Majlis}.
Throughout this study,
we work under the assumption that 
the N\'eel phase remains stable
in the presence of energy dissipation.
Elementary magnetic excitations are two kinds of magnons  
$\sigma=\pm$,
acoustic mode for $\sigma=+$
and optical mode for $\sigma=-$.

By incorporating 
the effect of energy dissipation on spins
into the energy dispersion relation
of magnons
through 
the two-coupled Landau-Lifshitz-Gilbert equation
where the value of the Gilbert damping constant 
$\alpha >0$
for each sublattice
is identical to each other,
we study the low-energy magnon dynamics~\cite{NHSM}
described by the energy dispersion relation
$\epsilon_{\sigma, {\mathbf{k}}, \alpha} \in {\mathbb{C}}$
of
$ {\text{Re}} 
(\epsilon_{\sigma, {\mathbf{k}}, \alpha}) \geq 0$
and the wavenumber 
${\mathbf{k}}=(k_x, k_y, k_z) \in {\mathbb{R}} $~\footnote{
As an example, 
Ref.~\cite{Superluminal} assumes
$\epsilon_{\sigma, {\mathbf{k}}, \alpha} \in {\mathbb{R}}$
and ${\mathbf{k}} \in {\mathbb{C}} $,
which describes
a spatially-decaying solution~\cite{NHrepresentation}.}
in the long wavelength limit
as~\cite{Superluminal}
\begin{equation}
\epsilon_{\sigma, {\mathbf{k}}, \alpha}
=
\frac{2S}{1+\alpha^2}
\Big(-i\alpha C
+\sqrt{(E_{\sigma, {\mathbf{k}}, \alpha})^2}\Big)   
\label{eqn:dispersion}
\end{equation}
and
\begin{equation}
(E_{\sigma, {\mathbf{k}}, \alpha})^2
:=
{A_{\sigma, \alpha}}^2 (a k)^2 
+{\delta_{\sigma}}^2 
-{{\cal{D}}_{\sigma}}^2 {\alpha}^2, 
\label{eqn:E}
\end{equation}
where 
$k:= |{\mathbf{k}} | $,
the length of a magnetic unit cell is $a$,
the spin moment in a magnetic unit cell is $S$,
and 
the others are material-dependent parameters
which are independent of the wavenumber,
$ 0< {A_{\sigma, \alpha}} \in {\mathbb{R}}$,   
$ 0< {\delta_{\sigma}} \in {\mathbb{R}}$, 
$0< {\cal{D}}_{\sigma} \in {\mathbb{R}}$,
and $ 0< C \in {\mathbb{R}}$:
The parameters are given as~\cite{Superluminal} 
\begin{subequations}
\begin{align}
{A_{\sigma,\alpha}}
=& \sqrt{(1+\alpha^2) 
\Big(J^2 +\sigma \frac{K_{\text{h}}}{2} J\Big)}, 
\label{eqn:A} \\
{\delta_{\sigma}}
=&
\sqrt{K_{\text{e}}(2J+K_{\text{e}}) 
+K_{\text{h}}(J-\sigma J+K_{\text{e}})}, 
\label{eqn:smalldelta} \\
{\cal{D}}_{\sigma}
=&
\sqrt{J^2+\sigma K_{\text{h}} J+ \frac{{K_{\text{h}}}^2}{4}}, 
\label{eqn:D}  \\
C
=&
J+K_{\text{e}}+\frac{K_{\text{h}}}{2}.
\label{eqn:C}
\end{align}
\end{subequations}
In the absence of the hard-axis anisotropy $K_{\text{h}}=0$,
two kinds of magnons $\sigma =\pm$
are in degenerate states,
whereas the degeneracy is lifted by $K_{\text{h}}>0$.
Note that, in general, the effect of dipolar interactions is negligibly small in AFMs, and we neglect it throughout this study.

The Gilbert damping constant $\alpha $
is a dimensionless constant,
and
the energy dissipation rate increases as 
the Gilbert damping constant grows.
In the dissipationless system~\cite{magnonCasimir_KK},
the Gilbert damping constant is zero 
$\alpha = 0$.
The dissipative system of $\alpha > 0$
described by Eq.~\eqref{eqn:dispersion}
can be regarded as a non-Hermitian system for magnons
in the sense that 
the energy dispersion takes a complex value.
Note that 
the constant term 
in Eq.~\eqref{eqn:dispersion}, 
$-i \alpha C$,
is independent of the wavenumber
and
just shifts 
the purely imaginary part of the magnon energy dispersion
$ \epsilon_{\sigma, {\mathbf{k}}, \alpha}$.
For this reason [Eq.~\eqref{eqn:CasE}],
the constant term, $-i \alpha C$,
is not relevant to the magnonic Casimir effect.
We then define the magnon energy gap 
of Eq.~\eqref{eqn:dispersion}
as
$ \Delta_{\sigma,\alpha} := 
{\text{Re}} (\epsilon_{\sigma, k=0, \alpha})  $,
i.e.,
\begin{equation}
\Delta_{\sigma,\alpha} 
=
\frac{2S}{1+{\alpha}^2}
{\text{Re}}
\Big(\sqrt{
(E_{\sigma, k=0, \alpha})^2
}\Big).
\label{eqn:gap}
\end{equation}

\subsection{Magnonic exceptional point}
\label{subsec:III}

\begin{figure*}[t]
\begin{tabular}{|c|c|c|c|}
    \hline  NiO
    &  Magnon energy dispersion 
          $ \epsilon_{\sigma,k, \alpha}$
    &  Re($E_{\text{Cas}}$)  and
            Re($C_{{\text{Cas}}}^{[b]}$)
    &  Im($E_{\text{Cas}}$)  and
            Im($C_{{\text{Cas}}}^{[b]}$) 
            \\  \hline 
\raisebox{18mm}{(i)}
& \begin{minipage}[t]{0.625\columnwidth}
      \centering
      \includegraphics[width=0.99\textwidth]{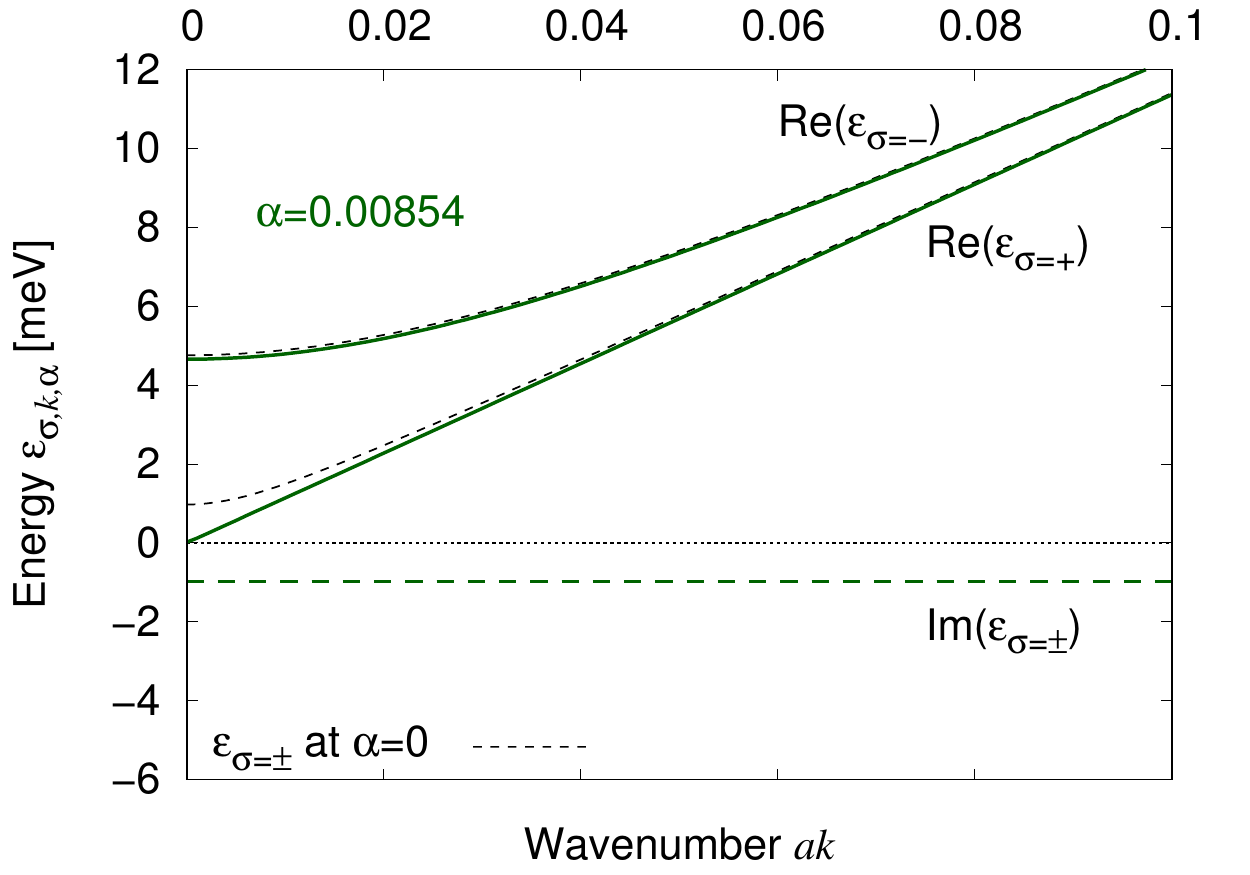}
    \end{minipage} &
    \begin{minipage}[t]{0.625\columnwidth}
      \centering
      \includegraphics[width=0.99\textwidth]{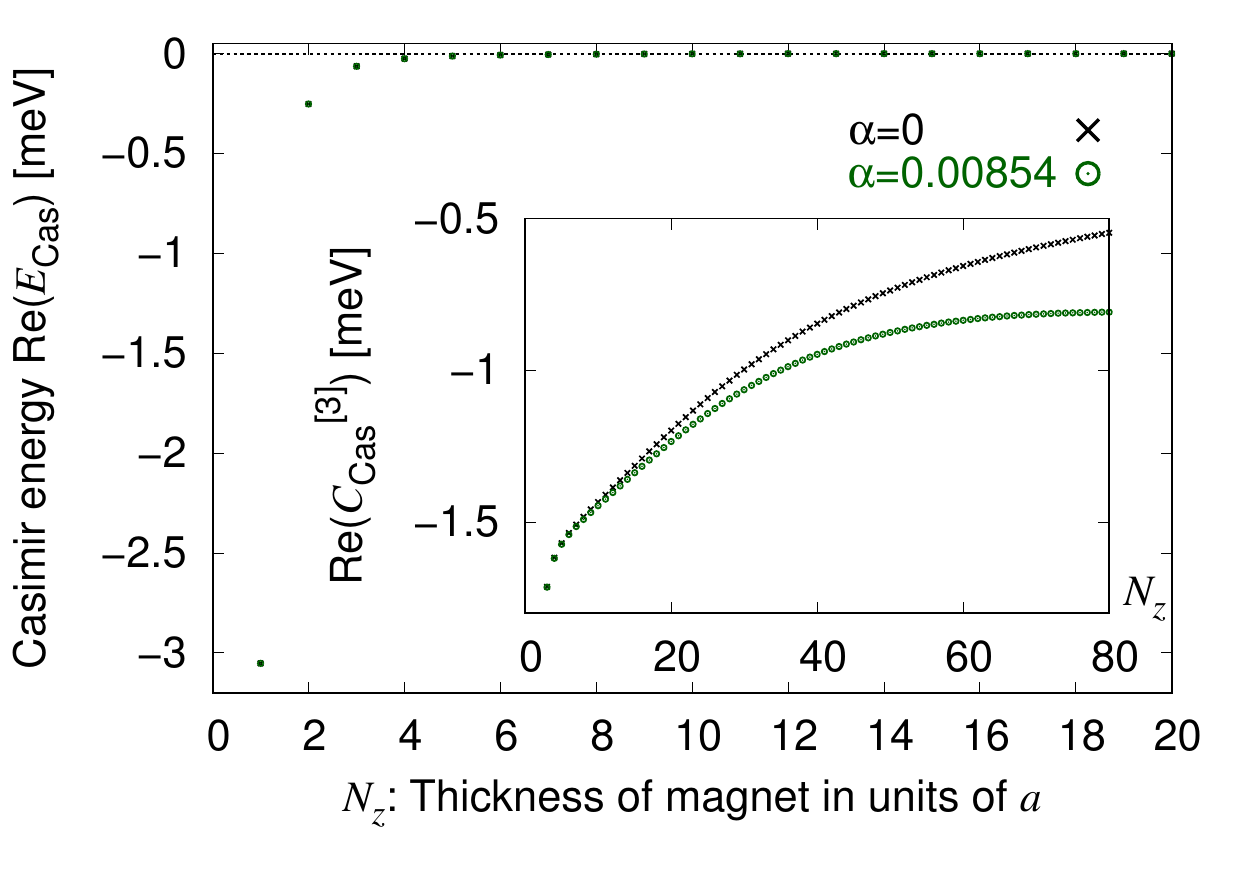}
    \end{minipage} &
    \begin{minipage}[t]{0.625\columnwidth}
      \centering
      \includegraphics[width=0.99\textwidth]{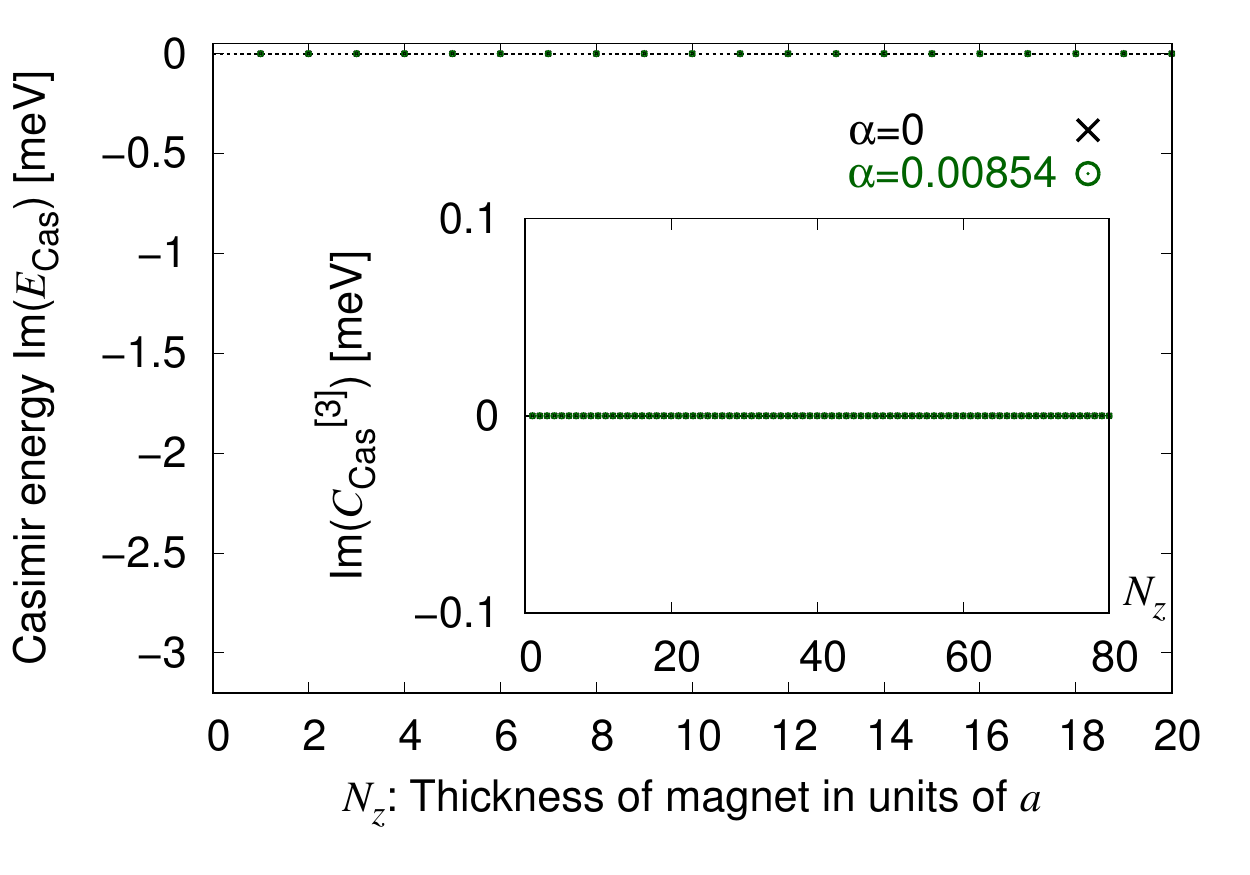}
    \end{minipage} 
    \\    \hline
\raisebox{18mm}{(ii)}
& \begin{minipage}[t]{0.625\columnwidth}
      \centering
      \includegraphics[width=0.99\textwidth]{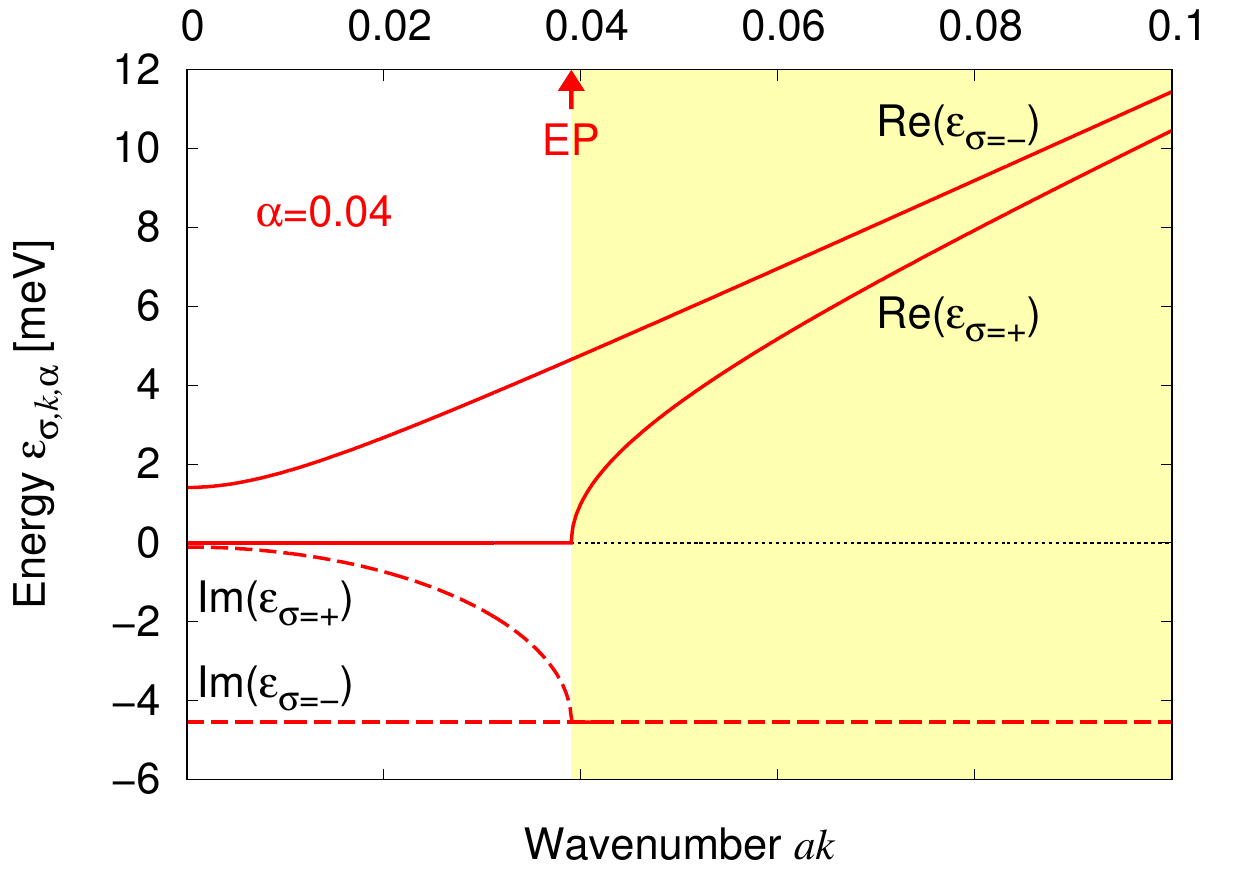}
    \end{minipage} &
    \begin{minipage}[t]{0.625\columnwidth}
      \centering
      \includegraphics[width=0.99\textwidth]{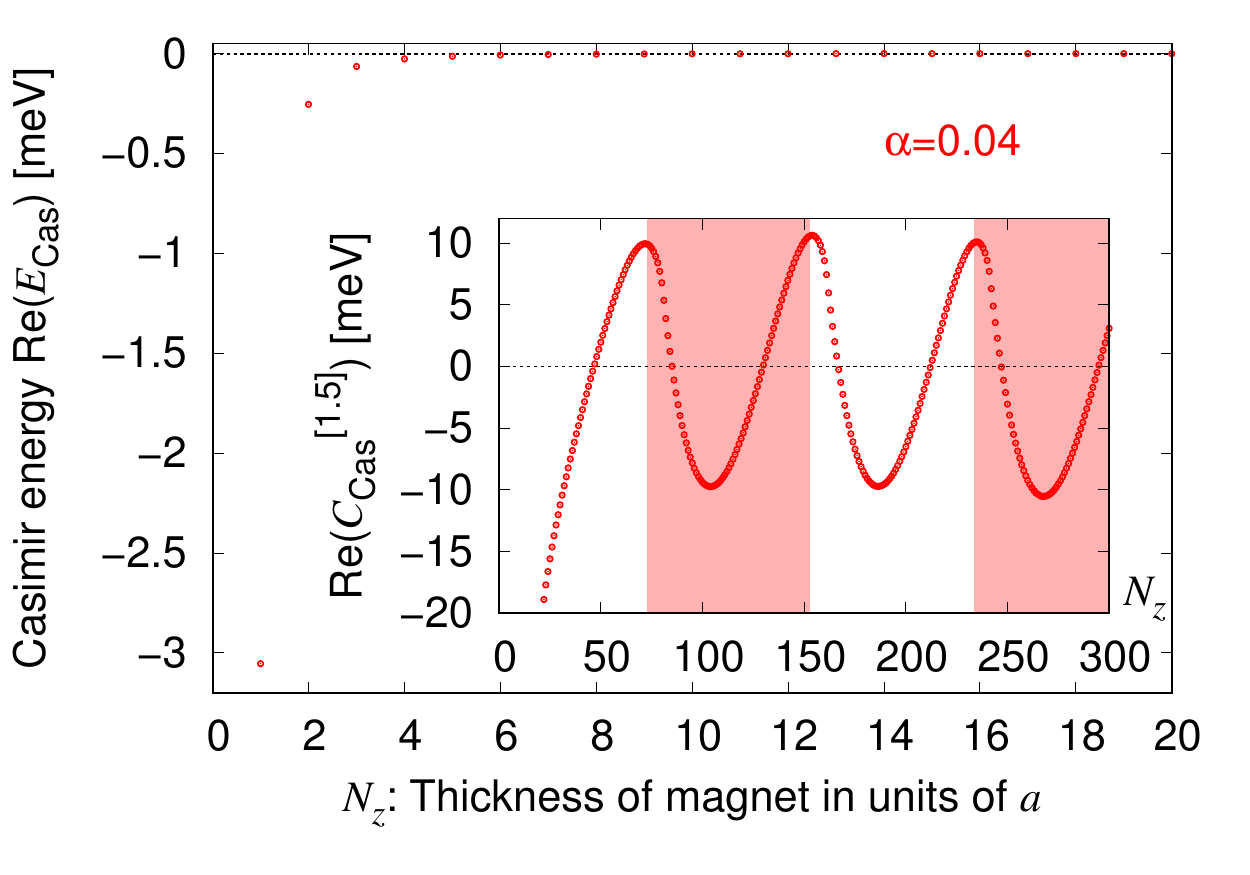}
    \end{minipage} &
    \begin{minipage}[t]{0.625\columnwidth}
      \centering
      \includegraphics[width=0.99\textwidth]{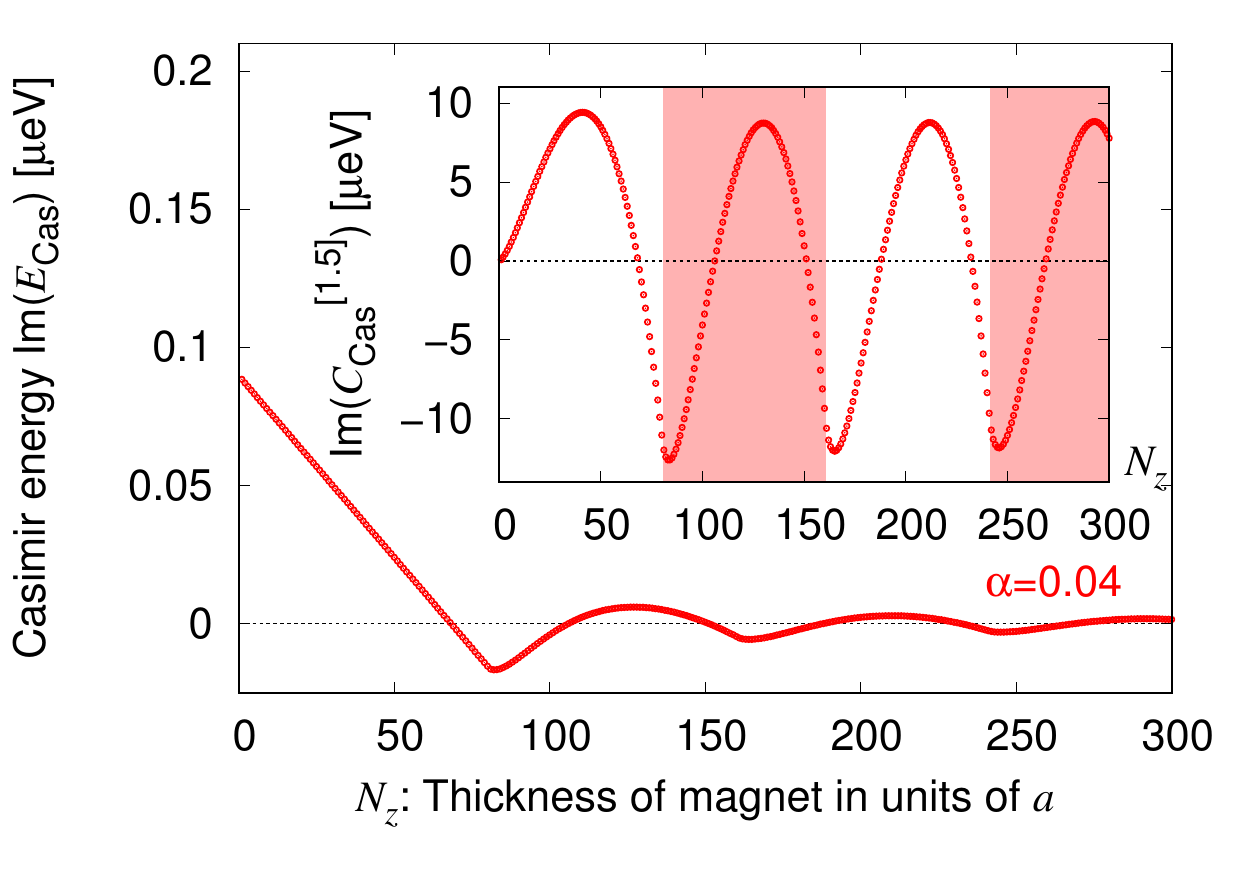}
    \end{minipage} \\    \hline
\raisebox{18mm}{(iii)}
& \begin{minipage}[t]{0.625\columnwidth}
      \centering
      \includegraphics[width=0.99\textwidth]{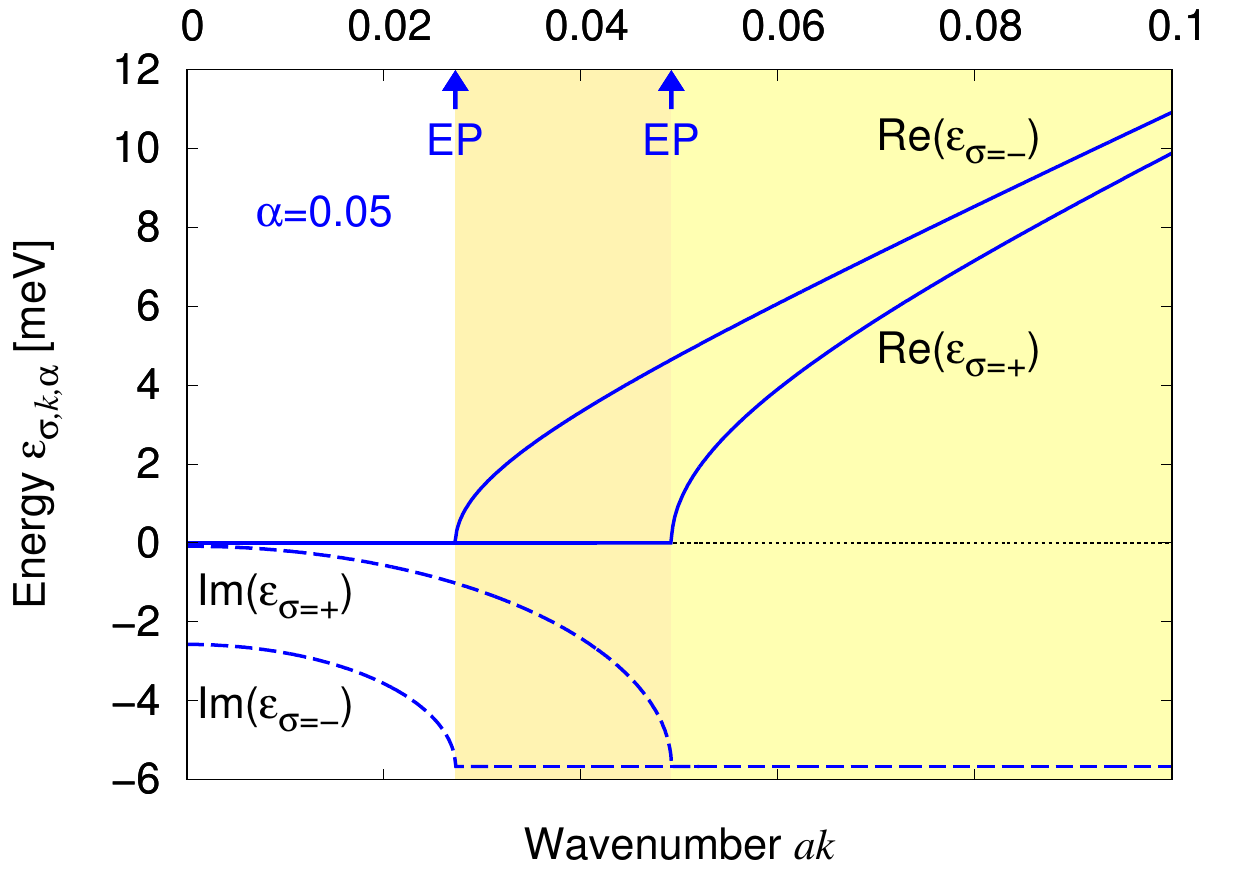}
    \end{minipage} &
    \begin{minipage}[t]{0.625\columnwidth}
      \centering
      \includegraphics[width=0.99\textwidth]{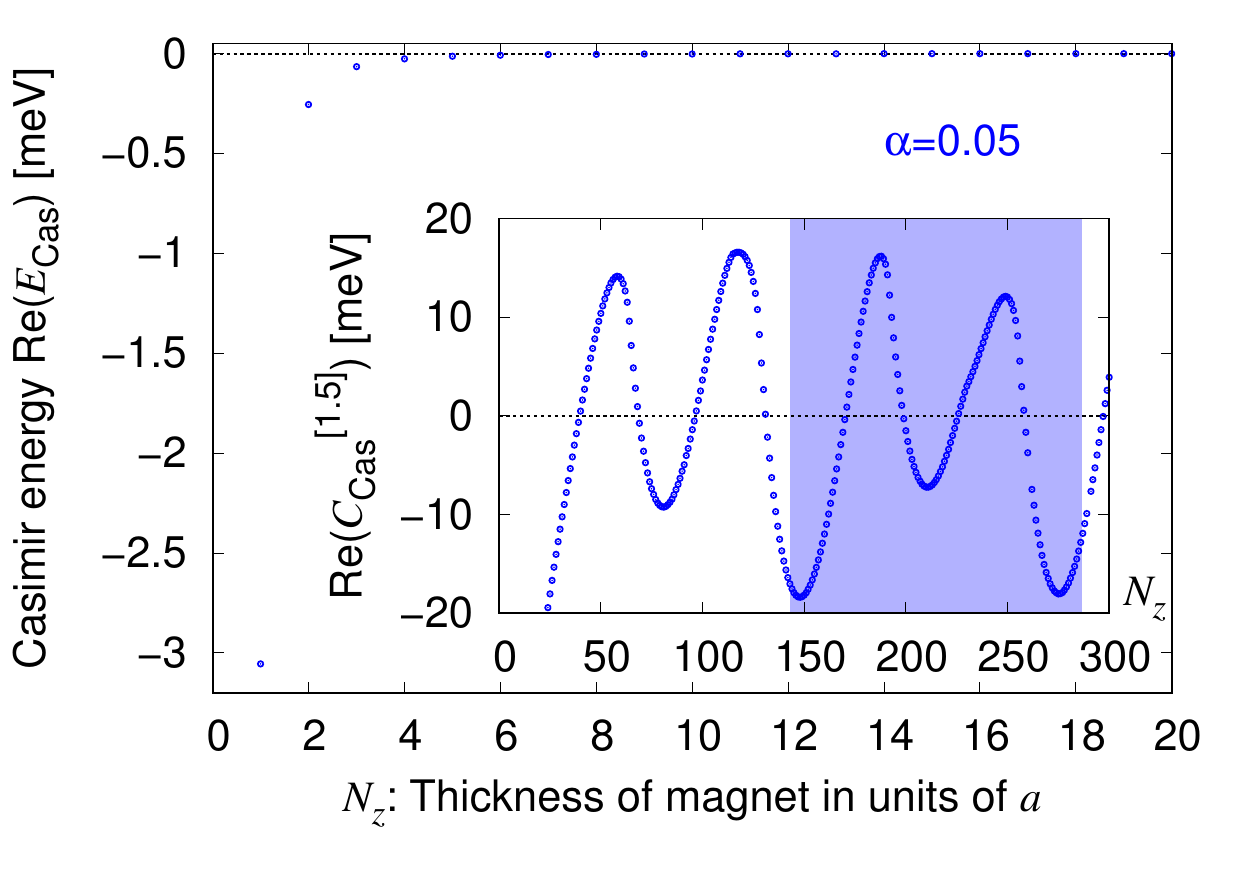}
    \end{minipage} &
    \begin{minipage}[t]{0.625\columnwidth}
      \centering
      \includegraphics[width=0.99\textwidth]{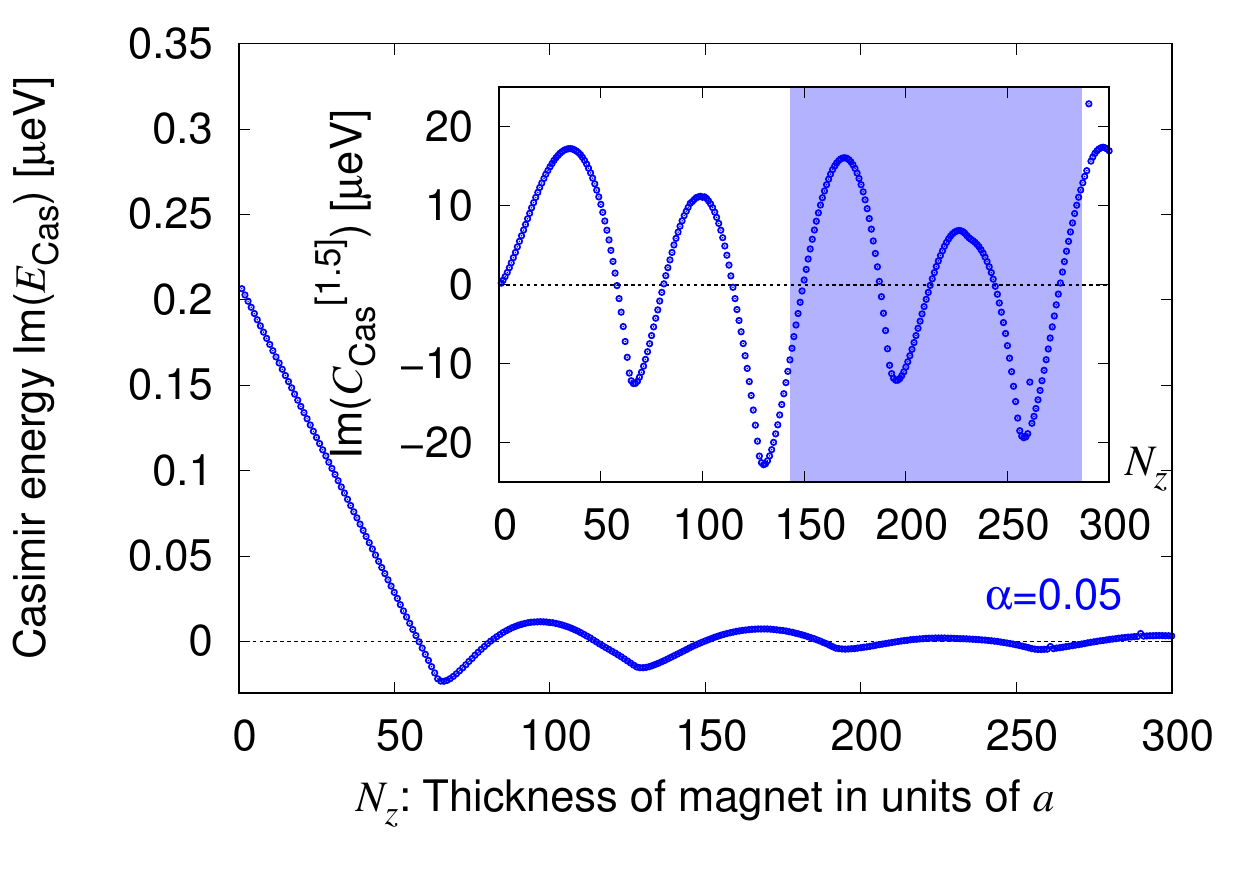}
    \end{minipage}
\\ \hline
\end{tabular}
\caption{Plots of
the magnon energy dispersion
$ \epsilon_{\sigma,k, \alpha}$,
the real part of the magnonic Casimir energy
$ \text{Re}(E_{\text{Cas}})$,
and
the imaginary part 
$ \text{Im}(E_{\text{Cas}})$
for NiO
in 
(i) 
the gap-melting regime,
(ii) 
the oscillating regime,
and
(iii) 
the beating regime.
Inset:
Each magnonic Casimir coefficient
$C_{{\text{Cas}}}^{[b]}$.
}
\label{fig:NiO}
\end{figure*}

When the damping constant $\alpha$ 
is small and
$(E_{\sigma, k=0, \alpha})^2 >0$,
$E_{\sigma, k=0, \alpha}$ takes a real value
and decreases as $\alpha$ increases.
This results in 
\begin{equation}
\frac{d \Delta_{\sigma,\alpha}}{d \alpha} 
< 0.
\label{eqn:gap_alpha}
\end{equation}
Thus, the magnon energy gap decreases 
as the damping constant increases~\cite{YT_EP_LLG}
[compare the solid line with the dashed one
in the left panel of Fig.~\ref{fig:NiO}~(i)].
When the damping constant is large enough,
the magnon energy gap vanishes
$ \Delta_{\sigma,\alpha} =0  $
at
$ \alpha =  {\alpha}_{\sigma}^{\text{cri}}  $,
\begin{equation}
 {\alpha}_{\sigma}^{\text{cri}}
:=
\frac{\delta_{\sigma}}{{{\cal{D}}_{\sigma}}},
\label{eqn:alpha_c}
\end{equation}
where
there exists the gapless magnon mode
which behaves like a relativistic particle 
with the linear energy dispersion.
From the property of
Eq.~\eqref{eqn:gap_alpha},
we call
(i)
$ \alpha \leq {\alpha}_{\sigma}^{\text{cri}}  $
the gap-melting regime.

When the damping constant exceeds the critical value
$ {\alpha}_{\sigma}^{\text{cri}}  $,
i.e., $ \alpha >   {\alpha}_{\sigma}^{\text{cri}} $,
$E_{\sigma, k=0, \alpha}$ takes a purely imaginary value
as $(E_{\sigma, k=0, \alpha})^2 <0$.
In this regime,
the real part of the magnon energy dispersion
remains zero
$ \text{Re} (\epsilon_{\sigma, {\mathbf{k}}, \alpha})=0  $
for the region
$0  \leq   k  \leq  k_{\sigma, \alpha}^{\text{cri}} $,
\begin{equation}
k_{\sigma, \alpha}^{\text{cri}}
:=\frac{1}{a}
\sqrt{\frac{ {{\cal{D}}_{\sigma}}^2 {\alpha}^2 - {\delta_{\sigma}}^2 }{{A_{\sigma,\alpha}}^2}},
\label{eqn:kc}
\end{equation}
whereas
$ \text{Re} (\epsilon_{\sigma, {\mathbf{k}}, \alpha})>0  $
for $ k  >  k_{\sigma, \alpha}^{\text{cri}} $
[see the highlighted in yellow
in the left panel of Figs.~\ref{fig:NiO}~(ii) and~(iii)].
The critical point 
$ k_{\sigma, \alpha}^{\text{cri}} $
can be regarded as 
the EP~\cite{YT_EP_LLG} 
for the wavenumber $k$,
and we refer to it as the magnonic EP.
As the value of the damping constant becomes larger,
that of the EP increases
\begin{equation}
\frac{d   k_{\sigma, \alpha}^{\text{cri}}}{d \alpha} >0.
\label{eqn:kc_alpha}
\end{equation}

At the EP
$ k = k_{\sigma, \alpha}^{\text{cri}} $,
the group velocity 
$ {\mathbf{v}}_{\sigma,{\mathbf{k}}, \alpha} := \text{Re}
[\partial \epsilon_{\sigma, {\mathbf{k}}, \alpha}/(\partial \hbar {\mathbf{k}})]$
becomes discontinuous 
[see the solid lines 
in the left panel of Figs.~\ref{fig:NiO}~(ii) and~(iii)].
In the vicinity of the EP,
the group velocity becomes much larger than the usual
such as in the gap-melting regime (i)
[compare the solid lines 
in the left panel of Figs.~\ref{fig:NiO}~(ii) and~(iii)
with the one of Fig.~\ref{fig:NiO}~(i)].

Assuming
${\alpha}_{\sigma=+}^{\text{cri}} 
< {\alpha}_{\sigma=-}^{\text{cri}} $,
the non-Hermitian system for magnons 
described by Eq.~\eqref{eqn:dispersion}
of $\alpha > 0$
can be divided into three regimes (i)-(iii)
in terms of the magnonic EPs
as follows 
[see the left panel of Figs.~\ref{fig:NiO}~(i),~(ii), and~(iii)]:

\begin{enumerate}[(i)]
\setlength{\parskip}{0.001cm}
\setlength{\itemsep}{0.01cm}
 \item 
 $ \alpha 
\leq {\alpha}_{\sigma=+}^{\text{cri}} 
< {\alpha}_{\sigma=-}^{\text{cri}} $.
No magnonic EPs.
 \item 
 $ {\alpha}_{\sigma=+}^{\text{cri}} 
< \alpha 
< {\alpha}_{\sigma=-}^{\text{cri}} $.
One EP,
$ k_{\sigma=+, \alpha}^{\text{cri}}  $.
 \item 
 $ {\alpha}_{\sigma=+}^{\text{cri}} 
< {\alpha}_{\sigma=-}^{\text{cri}} 
\leq \alpha $.
Two EPs, 
$ k_{\sigma=+, \alpha}^{\text{cri}}  $
and
$ k_{\sigma=-, \alpha}^{\text{cri}}  $.
\end{enumerate}


\subsection{Magnonic Casimir energy}
\label{subsec:IV}

The magnonic analog of the Casimir energy,
called the magnonic Casimir energy~\cite{magnonCasimir_KK},
is characterized by 
the energy dispersion relation of magnons.
Therefore,
by incorporating
the effect of energy dissipation on spins
into the energy dispersion relation of magnons
through the Gilbert damping constant 
[Eq.~\eqref{eqn:dispersion}],
a non-Hermitian extension 
of the magnonic Casimir effect
can be developed.
We remark that 
the Casimir energy induced by quantum fields on the lattice,
such as the magnonic Casimir energy~\cite{magnonCasimir_KK},
can be defined by using the lattice regularization~\cite{actor2000casimir,pawellek2013finite,CasimirKS3,CasimirKS4,KSremnantCasimir,mandlecha2022lattice,KSoscillation}.
In this study, 
we focus on thin films confined in the $z$ direction
(Fig.~\ref{fig:system}).
In the two-sublattice systems,
the wavenumber on the lattice is replaced as
$ ({a} k_j)^2 
\rightarrow 
2[1-\text{cos}({a} k_j)]$
along the $j$ axis for $j=x,y,z$.
Here by taking into account the Brillouin zone (BZ),
we set the boundary condition for the $z$ direction
in wavenumber space 
so that it is discretized as
$k_z \rightarrow \pi n/{L_z} $,
i.e., $ {a} k_z  \rightarrow \pi n/{N_z} $,
where 
$L_z := {a} N_z $ is the film thickness,
$ N_{j} \in {\mathbb{N}}$ is the number of magnetic unit cells 
along the $j$ axis,
and $n=1,2,..., 2N_z$.
Thus,
the magnonic Casimir energy
$ E_{\text{Cas}} $~\cite{magnonCasimir_KK} 
per the number of magnetic unit cells
on the surface for $N_z$
is defined 
as the difference 
between 
the zero-point energy
$ E_0^{\text{sum}} $ 
for the discrete energy 
$  \epsilon_{\sigma,{\mathbf{k}},\alpha ,n} $
due to discrete $k_z$
[Eq.~\eqref{eqn:CasEdisc}]
and the one
$ E_0^{\text{int}} $ 
for the continuous energy 
$ \epsilon_{\sigma,{\mathbf{k}}, \alpha} $
[Eqs.~\eqref{eqn:CasEcont} and~\eqref{eqn:dispersion}]
as follows~\cite{actor2000casimir,pawellek2013finite,CasimirKS3,CasimirKS4,KSremnantCasimir,mandlecha2022lattice,KSoscillation}:
\begin{subequations}
\begin{align}
 E_{\text{Cas}} (N_z)
 :=& E_0^{\text{sum}}(N_z)- E_0^{\text{int}}(N_z),  
\label{eqn:CasE} \\
 E_0^{\text{sum}}(N_z)
:=& \sum_{\sigma=\pm}
\int_{\text{BZ}}
\frac{d^2 ({a} k_{\perp})}{(2\pi)^2}
\Bigg[
\frac{1}{2}
\Big(
\frac{1}{2}
\sum_{n=1}^{2N_z}
\epsilon_{\sigma,{\mathbf{k}},\alpha,n}
\Big)
\Bigg], 
\label{eqn:CasEdisc} \\
 E_0^{\text{int}}(N_z)
:=&
\sum_{\sigma=\pm}
\int_{\text{BZ}}
\frac{d^2 ({a} k_{\perp})}{(2\pi)^2}
\Bigg[
\frac{1}{2}
N_z
\int_{\text{BZ}}
\frac{d ({a} k_z)}{2\pi}
\epsilon_{\sigma,{\mathbf{k}},\alpha}
\Bigg],
  \label{eqn:CasEcont}
\end{align}
\end{subequations}
where
$ k_{\perp}:=\sqrt{{k_x}^2+{k_y}^2} $,
$ d^2 ({a} k_{\perp})=d({a} k_x) d({a} k_y) $,
the integral is over the first BZ,
and
the factor $1/2$ 
in Eqs.~\eqref{eqn:CasEdisc} and~\eqref{eqn:CasEcont}
arises from the zero-point energy of the scalar field.

We remark that~\cite{NHSM} 
assuming thin films of $ N_z \ll N_x, N_y $
(Fig.~\ref{fig:system}),
the zero-point energy in the thin film of
the thickness $N_z$ is
$ E_0^{\text{sum}}(N_z) N_x N_y $
and consists of two parts as 
$E_0^{\text{sum}}(N_z) =
 E_{\text{Cas}} (N_z)
 + E_0^{\text{int}}(N_z)$
[Eq.~\eqref{eqn:CasE}],
where 
$E_0^{\text{int}}(N_z) $ exhibits 
the behavior of
$E_0^{\text{int}}(N_z) \propto  N_z $
[Eq.~\eqref{eqn:CasEcont}].
Then, to see the film thickness dependence of 
$ E_{\text{Cas}} (N_z)$,
we introduce 
the rescaled Casimir energy 
$ C_{{\text{Cas}}}^{[b]} $
in terms of ${N_z}^{b}  $ for $ b \in {\mathbb{R}}$ as
\begin{equation}
C_{{\text{Cas}}}^{[b]}(N_z):= 
 E_{\text{Cas}} \times  {N_z}^{b}
  \label{eqn:CasC}
\end{equation}
and call $ C_{{\text{Cas}}}^{[b]} $
the magnonic Casimir coefficient in the sense that
$E_{\text{Cas}} = C_{{\text{Cas}}}^{[b]} {N_z}^{-b}$.

Note that the zero-point energy arises from quantum fluctuations and does exist even at zero temperature. The zero-point energy defined at zero temperature does not depend on the Bose-distribution function 
[Eqs.~\eqref{eqn:CasEdisc} and~\eqref{eqn:CasEcont}]. Throughout this work, we focus on zero temperature~\cite{NHSM}.

\subsection{Magnonic non-Hermitian  
Casimir effect}
\label{subsec:V}

Finally, 
we investigate the magnonic Casimir effect 
in the non-Hermitian system $\alpha > 0$,
which we call the magnonic non-Hermitian Casimir effect, 
for each regime (i)-(iii).
As an example,
we consider NiO, an insulating AFM.
From Refs.~\cite{Superluminal,NiO1983,NiO2017},
we roughly estimate the model parameter values 
for NiO
as follows [Eq.~\eqref{eqn:dispersion}]:
$J=47.1$ meV,
$K_{\text{h}}=0.039\,5$ meV,
$K_{\text{e}}=0.001\,72 $ meV,
$S=1.21$,
and
$a=0.417$ nm.
NiO is a biaxial AFM of
$ K_{\text{h}} > 0$
and 
$ K_{\text{e}} > 0 $.
Due to the hard-axis anisotropy 
$K_{\text{h}} > 0$,
the degeneracy between two kinds of magnons $\sigma =\pm$
is lifted in NiO.
These parameters provide
$ {\alpha}_{\sigma=+}^{\text{cri}}\sim 0.008\,54
< {\alpha}_{\sigma=-}^{\text{cri}}\sim 0.041\,9 $.
Figure~\ref{fig:NiO} 
shows 
the magnon energy dispersion
[Eq.~\eqref{eqn:dispersion}]
and
the magnonic Casimir energy
[Eq.~\eqref{eqn:CasE}]
with its Casimir coefficient
[Eq.~\eqref{eqn:CasC}]
for each regime (i)-(iii).

\subsubsection{Gap-melting regime}
\label{subsubsec:VA}

(i) Gap-melting regime
$ \alpha 
\leq {\alpha}_{\sigma=+}^{\text{cri}} 
< {\alpha}_{\sigma=-}^{\text{cri}} $.
The magnonic Casimir energy 
takes a real value 
as shown in
the middle and right panels of
Fig.~\ref{fig:NiO}~(i),
see also Eq.~\eqref{eqn:CasC},
and there are no magnonic EPs
[see the left panel of Fig.~\ref{fig:NiO}~(i)].

When 
$ \alpha < {\alpha}_{\sigma=+}^{\text{cri}} $,
the magnon energy gap for both $  \sigma=\pm $ 
is nonzero
$ \Delta_{\sigma=\pm,\alpha} >0 $
and
both magnons $  \sigma=\pm $ are the gapped modes.
For each gapped mode,
the absolute value of the magnonic Casimir coefficient
$  C_{{\text{Cas}}}^{[3]} $
decreases and approaches asymptotically to zero 
as the film thickness increases.
We emphasize that 
the magnon energy gap decreases 
as the damping constant $\alpha$ increases
[Eq.~\eqref{eqn:gap_alpha}].
Then, the magnitude of the magnonic Casimir energy
and its coefficient increase
as the value of the damping constant becomes larger
and approaches to the critical value 
$ \alpha \rightarrow {\alpha}_{\sigma=+}^{\text{cri}} $
[see the middle panel of Fig.~\ref{fig:NiO}~(i)].

When 
$ \alpha = {\alpha}_{\sigma=+}^{\text{cri}} $,
the magnon $  \sigma=- $ remains the gapped mode,
whereas
the magnon energy gap for $  \sigma=+ $ vanishes
$ \Delta_{\sigma=+,\alpha} =0 $
and
the magnon $  \sigma=+ $ becomes 
the gapless mode
which behaves like a relativistic particle 
with the linear energy dispersion.
In the gapless mode,
the magnonic Casimir coefficient
$  C_{{\text{Cas}}}^{[3]} $
approaches asymptotically to a nonzero constant
as the film thickness increases.
The behavior of the gapless magnon mode is analogous to the conventional Casimir effect of a massless scalar field in continuous space~\cite{ambjorn1983properties} except for $a$-dependent lattice effects, 
whereas that of the gapped magnon modes is similar to the Casimir effect known for massive degrees of freedom~\cite{hays1979vacuum,ambjorn1983properties}.

\subsubsection{Oscillating regime}
\label{subsubsec:VB}

(ii) Oscillating regime 
$ {\alpha}_{\sigma=+}^{\text{cri}} 
< \alpha 
< {\alpha}_{\sigma=-}^{\text{cri}} $.
The magnonic Casimir energy 
takes a complex value 
as shown in the middle and right panels of
Fig.~\ref{fig:NiO}~(ii),
see also Eq.~\eqref{eqn:CasC}.
There is one EP,
e.g.,
$  a k_{\sigma=+, \alpha=0.04}^{\text{cri}} 
\sim  0.039\,1 $
for
$\alpha=0.04  $
[see the left panel of Fig.~\ref{fig:NiO}~(ii)].
Then, the magnonic non-Hermitian Casimir effect
exhibits an oscillating behavior
as a function of $N_z$
for the film thickness $L_z := {a} N_z $.

An intuitive explanation for 
the oscillation of the magnonic non-Hermitian Casimir effect
and its relation to the EP
is given as follows:
Through the lattice regularization,
the magnonic Casimir energy 
is defined as the difference 
[Eq.~\eqref{eqn:CasE}]
between 
the zero-point energy with the discrete wavenumber $k_z$
[Eq.~\eqref{eqn:CasEdisc}]
and the one with the continuous wavenumber
[Eq.~\eqref{eqn:CasEcont}].
On the lattice,
the wavenumber $k_z$ 
under the boundary condition
is discretized in units of 
$ \pi/a N_z $
as
$ k_z \rightarrow (\pi /a{N_z})n $.
As the film thickness $ N_z $ increases,
the unit becomes smaller,
and finally, it matches the EP as
$ \pi/a N_z = k_{\sigma, \alpha}^{\text{cri}}$,
i.e., $ N_z = \pi/a k_{\sigma, \alpha}^{\text{cri}}  $,
where the magnonic non-Hermitian Casimir effect is enhanced
due to the EP.
Then,
the magnonic non-Hermitian Casimir effect is
periodically enhanced 
where the film thickness $ N_z $ is multiples of 
$ \pi/a k_{\sigma, \alpha}^{\text{cri}}$.
Thus, 
the oscillating behavior of 
the magnonic non-Hermitian Casimir effect 
stems from the EP,
$ k_{\sigma, \alpha}^{\text{cri}} $,
and the oscillation is characterized in units of
$  {\pi}/a {k_{\sigma, \alpha}^{\text{cri}}} $.
We refer to this oscillating behavior as
the magnonic EP-induced Casimir oscillation.
The period of this Casimir oscillation is 
\begin{equation}
{\Lambda}_{\sigma, \alpha}^{\text{Cas}} 
:=\frac{\pi}{a k_{\sigma, \alpha}^{\text{cri}}}.
  \label{eqn:PeriodOscillation}
\end{equation}
As an example,
the period is
$  {\Lambda}_{\sigma=+, \alpha=0.04}^{\text{Cas}}
\sim 80.4  $
for $\alpha=0.04  $.
This agrees with the numerical result
in the middle and right panels
of Fig.~\ref{fig:NiO}~(ii),
see the highlighted in red.
We call
(ii)
$ {\alpha}_{\sigma=+}^{\text{cri}} 
< \alpha 
< {\alpha}_{\sigma=-}^{\text{cri}} $
the oscillating regime.
The middle and right panels of
Fig.~\ref{fig:NiO}~(ii) show that
the magnonic EP-induced Casimir oscillation 
is characterized by its Casimir coefficient $ C_{{\text{Cas}}}^{[b]} $ of $b=1.5$.

\subsubsection{Beating regime}
\label{subsubsec:VC}

(iii) Beating regime
$ {\alpha}_{\sigma=+}^{\text{cri}} 
< {\alpha}_{\sigma=-}^{\text{cri}} 
\leq \alpha $.
The magnonic Casimir energy 
takes a complex value 
as shown in the middle and right panels of
Fig.~\ref{fig:NiO}~(iii),
see also Eq.~\eqref{eqn:CasC}.
There are two EPs, 
$ k_{\sigma=+, \alpha}^{\text{cri}}  $
and
$ k_{\sigma=-, \alpha}^{\text{cri}} $,
which induce
two types of the Casimir oscillations
characterized by 
$ {\Lambda}_{\sigma=+, \alpha}^{\text{Cas}} $
and
$ {\Lambda}_{\sigma=-, \alpha}^{\text{Cas}} $,
respectively.
As an example,
$  a k_{\sigma=+, \alpha=0.05}^{\text{cri}} 
\sim  0.049\,2 $
and
$  a k_{\sigma=-, \alpha=0.05}^{\text{cri}} 
\sim  0.027\,3 $
provide
$  {\Lambda}_{\sigma=+, \alpha=0.05}^{\text{Cas}}
\sim  63.8 $
and
$  {\Lambda}_{\sigma=-, \alpha=0.05}^{\text{Cas}}
\sim  115 $,
respectively,
for $\alpha=0.05$
[see the left panel of Fig.~\ref{fig:NiO}~(iii)].
Due to the interference between the two Casimir oscillations,
the magnonic non-Hermitian Casimir effect 
exhibits a beating behavior
as a function of $N_z$
for the film thickness $L_z := {a} N_z $
with a period of 
\begin{equation}
\frac{1}{| 1/{\Lambda}_{\sigma=+, \alpha}^{\text{Cas}}
-1/{\Lambda}_{\sigma=-, \alpha}^{\text{Cas}}|}.
  \label{eqn:PeriodOscillationBeat}
\end{equation}
As an example,
the period is
$| 1/{\Lambda}_{\sigma=+, \alpha=0.05}^{\text{Cas}}
-1/{\Lambda}_{\sigma=-, \alpha=0.05}^{\text{Cas}}|^{-1}
\sim 143$
for $ \alpha=0.05 $.
This agrees with the numerical result
in the middle and right panels of
Fig.~\ref{fig:NiO}~(iii),
see the highlighted in blue.
We call 
(iii) 
$ {\alpha}_{\sigma=+}^{\text{cri}} 
< {\alpha}_{\sigma=-}^{\text{cri}} 
\leq \alpha $
the beating regime.
The middle and right panels of
Fig.~\ref{fig:NiO}~(iii)
show that
the beating behavior of 
the magnonic EP-induced Casimir oscillation 
is characterized by its Casimir coefficient $ C_{{\text{Cas}}}^{[b]} $ of $b=1.5$.
We remark that
the beating behavior is absent
in the uniaxial AFMs of $K_{\text{h}}=0$
and
$K_{\text{e}}>0$
where
two kinds of magnons $\sigma =\pm$
are in degenerate states~\cite{NHSM}.


\subsubsection{Imaginary part of the Casimir energy}
\label{subsubsec:VD}

Here,
we discuss the meaning of the imaginary part of the Casimir energy.
The (complex) Casimir energy is defined by the zero-point energy which is the sum of all the possible (complex) eigenvalues. 
The real part of the zero-point energy originates from the sum of the real parts of the eigenvalues, 
whereas the imaginary part of the zero-point energy is defined as the sum of imaginary parts of eigenvalues. 
Since the imaginary parts of eigenvalues are formally regarded as the decay width (or the inverse of a lifetime) of an unstable particle, 
the imaginary part of the zero-point energy is the sum of all the possible decay widths.
Hence, 
if the decay width of an unstable particle 
depends on the wavenumber, 
and the width in the thin film and that in the bulk 
are different from each other, 
then the imaginary part of the Casimir energy can be nonzero.
In this work, 
since we focus on magnons
in the geometry of Fig.~\ref{fig:system}, 
the imaginary part of magnonic Casimir energy 
represents the $L_z$-dependence of the sum of magnon decay widths.

\section{Discussion}
\label{sec:VI}

\subsection{Magnonic Casimir engineering}
\label{subsec:VI}

The Gilbert damping can be enhanced
and controlled 
by the established experimental techniques of spintronics
such as
spin pumping~\cite{NHSM}.
In addition,
microfabrication technology
can control the film thickness
and manipulate the magnonic non-Hermitian Casimir effect.
The Casimir pressure of magnons,
which stems from
the real part of its Casimir energy,
contributes to 
the internal pressure of thin films.
We find from
the middle panel of
Figs.~\ref{fig:NiO}~(ii) and~(iii)
that depending on the film thickness,
the sign of 
the real part of the magnonic Casimir coefficient 
changes.
This means that by tuning the film thickness,
we can control and manipulate the direction of the magnonic Casimir pressure as well as the magnitude
thanks to
the EP-induced Casimir oscillation.
Thus, our study utilizing energy dissipation,
the magnonic non-Hermitian Casimir effect,
provides the new principles of nanoscale devices, 
such as highly sensitive pressure sensors and magnon transistors~\cite{MagnonTransistor},
and paves a way for magnonic Casimir engineering.

\subsection{Conclusion}
\label{subsec:VII}

We have shown that 
as the Gilbert damping constant increases,
the non-Hermitian Casimir effect of magnons
in antiferromagnets
is enhanced 
and exhibits the oscillating behavior
which stems from the exceptional point.
This exceptional point-induced Casimir oscillation 
also exhibits the beating behavior
when the degeneracy between two kinds of magnons
is lifted.
These magnonic Casimir oscillations
are absent in the dissipationless system of magnons.
Thus,
we have shown that 
energy dissipation
serves as a new handle on Casimir engineering.

\subsection{Outlook}
\label{subsec:VIII}

In this paper following Ref.~\cite{Superluminal},
the effect of dissipation 
is incorporated into the energy dispersion relation of magnons 
through the Landau-Lifshitz-Gilbert equation.
It will be intriguing to find the quantum effect of dissipation
on magnonic non-Hermitian Casimir effect, 
beyond the Landau-Lifshitz-Gilbert equation,
by using quantum master equation~\cite{ReviewNHmagnonics,QME2022Ji,QME2022Yuan,QME2024Ji}.
We also remark that 
dipolar interactions contribute to 
the form of the dispersion relation~\cite{YIGthin2008}
and play a crucial role in magnonic Casimir effect 
in ferrimagnets~\cite{magnonCasimir_KK}.
Hence, taking dipolar interactions into account,
it will be interesting to develop this study, 
magnonic non-Hermitian Casimir effect in antiferromagnets, 
into ferrimagnets. 
We leave these advanced studies for future works.


\

\


\noindent{\textbf{Methods}}

\noindent{Numerical calculation was performed by using the software Wolfram Mathematica.}


\

\noindent{\textbf{Data Availability}}

\noindent{No datasets were generated or analysed during this work.}


\

\noindent{\textbf{Acknowledgements}}

\noindent{We would like to thank
Ryo Hanai,
Hosho Katsura,
Norio Kawakami,
Se Kwon Kim,
Katsumasa Nakayama,
Masatoshi Sato,
Kenji Shimomura,
Ken Shiozaki,
Keisuke Totsuka,
Shun Uchino,
and
Hikaru Watanabe
for helpful comments and discussions.
We acknowledge support
by JSPS KAKENHI Grants 
No. JP20K14420 (K. N.), 
No. JP22K03519 (K. N.),
No. JP17K14277 (K. S.), 
and No. JP20K14476 (K. S.).}


\

\noindent{\textbf{Author Contributions}}

\noindent{The two authors contributed equally to this work.}


\

\noindent{\textbf{Competing Interests}}

\noindent{The authors declare no competing interests.}

\bibliography{PumpingRef}

\newpage

\appendix


\begin{center}
\noindent{\large{\textbf{Supplemental Material}}}
\end{center}

\renewcommand{\thesection}{S-\Roman{section}}

In this Supplemental Material,
we add an explanation about the Casimir energy 
induced by quantum fields on the lattice
and provide some details about 
the magnonic Casimir effect 
in the absence of the hard-axis anisotropy. 
We also add remarks on, in order, 
observation of the magnonic Casimir effect in the AFMs, 
Casimir effects from other origins,
thermal effects, 
higher energy bands, 
edge or surface magnon modes,
and the effect of the edge condition.

{\section{The Casimir energy on the lattice}
\label{sec:EcasLattice}
}

In the main text,
following the Casimir energy for photon fields
(i.e., quantum fields in continuous space)~\cite{CasimirEffect},
the magnonic Casimir energy
is defined 
as in Eqs.~\eqref{eqn:CasE},~\eqref{eqn:CasEdisc},
and~\eqref{eqn:CasEcont}
through the lattice regularization.
In contrast to the Casimir effect 
for photon fields
(i.e., quantum fields in continuous space),
the magnonic Casimir energy 
is induced by its quantum field on the lattice,
and there is no ultraviolet divergence
in each component
[see Eqs.~\eqref{eqn:CasEdisc} and~\eqref{eqn:CasEcont}].
Here we remark that 
the Casimir energy induced by quantum fields on the lattice,
such as the magnonic Casimir energy
$ E_{\text{Cas}} (N_z) $
[see Eq.~\eqref{eqn:CasE}],
plays a key role in finding 
the film thickness dependence of
the zero-point energy in the thin film
(see Fig.~\ref{fig:system}).
The zero-point energy in the thin film of
the thickness $N_z$ is
$ E_0^{\text{sum}}(N_z) N_x N_y $
[see Eq.~\eqref{eqn:CasEdisc}]
and consists of two parts as 
$E_0^{\text{sum}}(N_z) =
 E_{\text{Cas}} (N_z)
 + E_0^{\text{int}}(N_z)$
[see Eq.~\eqref{eqn:CasE}],
where 
$E_0^{\text{int}}(N_z) $ exhibits 
the behavior of
$E_0^{\text{int}}(N_z) \propto  N_z $
[see Eq.~\eqref{eqn:CasEcont}].


\

\

\section{The hard-axis anisotropy}
\label{sec:Khard}

\subsection{In the absence of the hard-axis anisotropy}
\label{subsec:KhardZero}

In the main text, 
we have considered NiO.
NiO is a biaxial AFM of
$K_{\text{h}}>0$
and $K_{\text{e}}>0$:
There exist
not only
the easy-axis anisotropy
$K_{\text{e}}=0.001\,71829 $ meV
but also 
the hard-axis anisotropy
$K_{\text{h}}=0.039\,5212$ meV,
see the main text for other parameter values.
Here,
by changing 
only the value of $K_{\text{h}} $
to $K_{\text{h}} = 0$
with leaving other parameter values unchanged,
we estimate
the magnonic Casimir effect 
and provide some details about its behavior
in the absence of the hard-axis anisotropy.

Figure~\ref{fig:SM_E} shows
the magnon energy dispersion
$ \epsilon_{\sigma,k, \alpha}$
for the gap-melting regime (i)
in the absence of the hard-axis anisotropy
$K_{\text{h}}=0$.
Figure~\ref{fig:SM_Ccas} shows
the real part of the magnonic Casimir energy
$ \text{Re}(E_{\text{Cas}})$
for the gap-melting regime (i)
in the absence of the hard-axis anisotropy
$K_{\text{h}}=0$
and that in the presence of hard-axis anisotropy 
$K_{\text{h}}=0.039\,5212$ meV.
The latter is the same as 
the middle panel of Fig.~\ref{fig:NiO}~(i).

In the absence of the hard-axis anisotropy
$K_{\text{h}}=0$,
two kinds of magnons $\sigma =\pm$
are in degenerate states
[see Eq.~\eqref{eqn:dispersion}].
This results in
$ {\alpha}_{\sigma=+}^{\text{cri}}
= {\alpha}_{\sigma=-}^{\text{cri}}
= 0.008\,54$
[see Eq.~\eqref{eqn:alpha_c}].
When the damping constant reaches the critical value
$ \alpha 
= {\alpha}_{\sigma=+}^{\text{cri}}
= {\alpha}_{\sigma=-}^{\text{cri}}
= 0.008\,54$,
the magnon energy gaps for both $ \sigma=\pm $ vanish,
$ \Delta_{\sigma=\pm,\alpha} =0 $,
and
both magnons $ \sigma=\pm $ become
the gapless modes
which behave like relativistic particles
with the linear energy dispersion
(see the solid line in Fig.~\ref{fig:SM_E}).
Then, the magnonic Casimir coefficient
$ C_{{\text{Cas}}}^{[3]} $
asymptotically approaches to a nonzero constant
as the film thickness increases
(see Fig.~\ref{fig:SM_Ccas}),
which means that 
its Casimir energy exhibits the behavior of
$ E_{\text{Cas}} \propto 1/{N_z}^{3} $.
Figure~\ref{fig:SM_Ccas}
also shows that 
the magnitude of the magnonic Casimir energy
and its coefficient 
for $K_{\text{h}}=0$ 
become larger than that 
for $K_{\text{h}}=0.039\,5212$ meV.

\renewcommand{\thefigure}{S\arabic{figure}}
\setcounter{figure}{0}
\begin{figure}[t]
\centering
\includegraphics[width=0.45\textwidth]{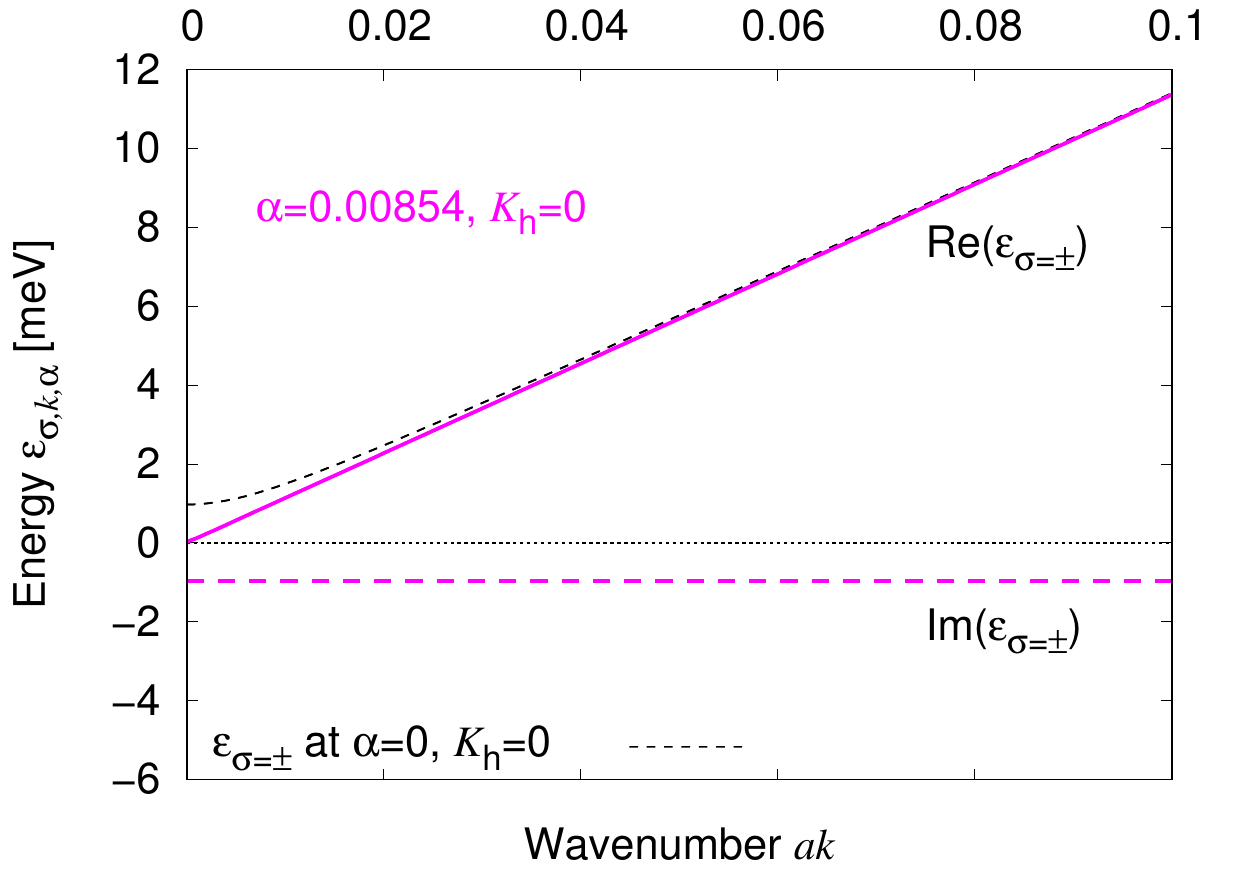}
\caption{Plots of
the magnon energy dispersion
$ \epsilon_{\sigma,k, \alpha}$
for the gap-melting regime (i)
in the absence of the hard-axis anisotropy
$K_{\text{h}}=0$,
where
$ {\alpha}_{\sigma=+}^{\text{cri}}
= {\alpha}_{\sigma=-}^{\text{cri}}
= 0.008\,54$.
}
\label{fig:SM_E}
\end{figure}

\begin{figure}[t]
\centering
\includegraphics[width=0.45\textwidth]{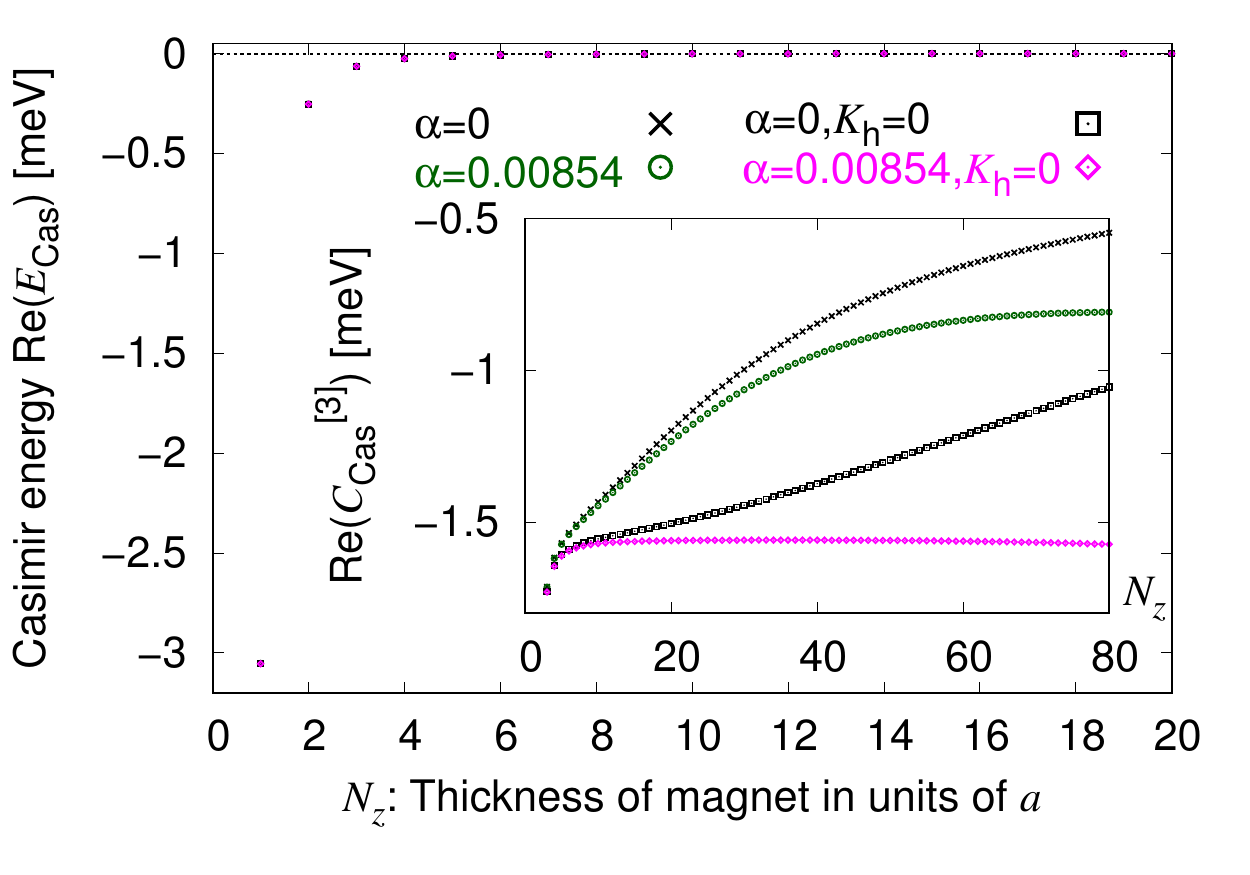}
\caption{Plots of 
the real part of the magnonic Casimir energy
$ \text{Re}(E_{\text{Cas}})$
for the gap-melting regime (i)
in the absence of the hard-axis anisotropy
$K_{\text{h}}=0$
and those
in the presence of hard-axis anisotropy 
$K_{\text{h}}=0.039\,5212$ meV.
The latter is the same as 
the middle panel of Fig.~\ref{fig:NiO}~(i).
Inset: Its Casimir coefficient 
$C_{{\text{Cas}}}^{[b]}
= E_{\text{Cas}} \times  {N_z}^{b}$.
}
\label{fig:SM_Ccas}
\end{figure}

\subsection{An example of the uniaxial AFM}
\label{subsec:uniAF}

As an example, 
$\text{Cr}_2\text{O}_3$
can be regarded as 
a uniaxial AFM of 
$K_{\text{h}}=0$
and $K_{\text{e}}>0$,
where
two kinds of magnons $\sigma =\pm$
are in degenerate states.
Hence, 
the magnonic EP-induced Casimir oscillation
is one type,
and its beating behavior is absent
in 
$\text{Cr}_2\text{O}_3$.
Magnonic non-Hermitian Casimir effects in the AFMs 
of $K_{\text{e}}>0$
are summarized in Table~\ref{tab:table1}.
Note that 
Ref.~\cite{THzMagnon2} reported
the experimental realization of
sub-terahertz spin pumping in 
$\text{Cr}_2\text{O}_3$,
and Ref.~\cite{AFpumpingTHz} 
reported that in NiO.

{\section{Remarks on observation}
\label{sec:Observation}
}

In the main text,
we have explained that 
the Gilbert damping can be enhanced
and controlled 
by the use of
the established experimental techniques of spintronics
such as
spin pumping.
Here we add remarks on 
observation of our theoretical prediction.
We expect that the magnonic Casimir effect in the AFMs
can be experimentally observed in principle 
through measurement of magnetization.
The reason is as follows.

External magnetic fields induce magnetostriction,
which can be regarded as a kind of lattice deformation,
and its correction for 
the length of a magnetic unit cell $a$
is characterized by the magnetostriction constant~\cite{belov1960magnetostriction,alberts1961magnetostriction,nakamichi1961antiferromagnetic,mcguire1962magnetic,smith1966magnetostriction,yamada1966magnetic,dudko1971magnetostriction,yacovitch1977magnetostriction}.
The magnonic Casimir energy of the AFMs does not depend on
external magnetic fields usually,
whereas the magnonic Casimir effect is influenced 
by magnetostriction,
and its correction for the magnonic Casimir energy 
depends on magnetic fields and contributes to magnetization.
Thus, although the correction is small,
the magnonic Casimir effect in the AFMs can be 
experimentally observed in principle 
through measurement of magnetization 
and its film thickness dependence
by using external magnetic fields 
(i.e., magnetostriction).

\renewcommand{\thetable}{T\arabic{table}}
\setcounter{table}{0}
\begin{table}
\caption{
\label{tab:table1}
Magnonic non-Hermitian Casimir effects 
in the AFMs
of $K_{\text{e}}>0$.
}
\begin{ruledtabular}
\begin{tabular}{lcc}

&  $K_{\text{h}}>0$
&  $K_{\text{h}}=0$      \\  \hline  
The degeneracy of magnons ($\sigma =\pm$)  &    
No  &  
Yes  \\  
The number of the magnonic EPs  & 
$2$  &  
$1$   \\
The EP-induced
Casimir oscillation & 
   Yes   &  
     Yes    \\
The beating behavior of the oscillation & 
 Yes  &  
  No
\end{tabular}
\end{ruledtabular}
\end{table}

We remark that the magnetic-field derivative 
of the real part of the Helmholtz free energy 
is magnetization.
At zero temperature,
assuming thin films of $ N_z \ll N_x, N_y $ 
(see Fig.~\ref{fig:system}),
the Helmholtz free energy of quantum fields for magnons 
in the thin film of the thickness $N_z$ 
is
$ E_0^{\text{sum}}(N_z) N_x N_y $
[see Eq.~\eqref{eqn:CasEdisc}]
and consists of two parts as 
$E_0^{\text{sum}}(N_z) =
 E_{\text{Cas}} (N_z)
 + E_0^{\text{int}}(N_z)$
[see Eq.~\eqref{eqn:CasE}],
where 
$E_0^{\text{int}}(N_z) $ exhibits 
the linear-in-$N_z$ behavior as
$E_0^{\text{int}}(N_z) \propto  N_z $
[see Eq.~\eqref{eqn:CasEcont}].
Since 
$  E_{\text{Cas}} (N_z)$
exhibits an oscillating and a beating behavior 
as a function of the film thickness 
in the regimes (ii) and (iii), respectively
[see Eq.~\eqref{eqn:CasC} 
and the middle panels of Figs.~\ref{fig:NiO}~(ii) and~(iii)],
the Helmholtz free energy of the thin film shows 
a different $ N_z $-dependence from 
the linear-in-$N_z$ behavior.
In other words,
magnetization of the thin film 
exhibits 
an oscillating or a beating behavior 
as a function of the film thickness
due to the magnonic non-Hermitian Casimir effect.
Hence, our prediction, 
the non-Hermitian Casimir effect of magnons,
can be observed in principle 
through measurement of magnetization,
its oscillating or beating behavior 
as a function of the film thickness.

{\section{Casimir effects from other origins
}
\label{sec:OtherOrigins}
}

In the main text,
we have focused on the magnonic Casimir effect.
Here we add a remark on Casimir effects from other origins 
such as phonons and photons. 
Even excluding magnetostriction,
the energy dispersion of magnons depends strongly on 
magnetic fields through Zeeman coupling, 
whereas those of phonons and photons do not. 
Therefore, we expect that 
the magnonic Casimir effect can be distinguished experimentally
from the others 
by manipulating external magnetic fields.

\

\

{\section{Thermal effects}
\label{sec:ThermalEffect}
}

In the main text, 
we have focused on zero temperature.
Here we remark on thermal effects.
At nonzero temperature, 
a thermal contribution to the Helmholtz free energy,
called the thermal Casimir energy,
arises additionally
and is characterized by the Boltzmann factor.
It should be emphasized that,
although it is called the thermal Casimir energy,
there is a significant distinction 
between  
the thermal Casimir effect
and
the Casimir effect:
The thermal Casimir effect is independent of
the zero-point energy.
The thermal Casimir effect arises from thermal fluctuations
and is affected by temperatures,
whereas
the Casimir effect arises from 
the zero-point energy
due to quantum fluctuations
and is not affected by temperatures.
Hence, we expect that 
the Casimir effect of magnons 
can be distinguished experimentally
from 
its thermal Casimir effect
by manipulating temperature.
For details of 
a magnonic analog of the thermal Casimir effect
in a Hermitian system,
see Ref.~\cite{RChengAFthermalCasimirMagnon}
as an example.

\

\

{\section{Higher energy bands}
\label{sec:HigherBand}
}

In the main text,
we have assumed that 
the magnonic Casimir energy of the AFM, NiO,
is dominated by
the two bands of Eq.~\eqref{eqn:dispersion}.
Here we remark on the contribution from higher energy bands
than those of Eq.~\eqref{eqn:dispersion}.
The magnonic Casimir energy or the zero-point energy 
[see Eq.~\eqref{eqn:CasE}]
arises from quantum fluctuations
and does exist even at zero temperature.
The zero-point energy defined at zero temperature
does not depend on the Bose-distribution function
[see Eqs.~\eqref{eqn:CasEdisc} and~\eqref{eqn:CasEcont}].
Hence, higher energy bands than 
those of Eq.~\eqref{eqn:dispersion}
also can contribute to the magnonic Casimir energy.
However, the contribution becomes smaller
as the shape of the bands is flatter.
Numerical calculations of Refs.~\cite{YIGdata3,ShamotoYIG,YIGspectrum2021}
show that 
higher energy bands of a ferrimagnet tend to be flat.
The ferrimagnet has an alternating structure of up and down spins
like the N\'eel magnetic order of the AFM,
and in this sense,
the ferrimagnet is similar to the AFM.
We therefore assume that 
higher energy bands of the AFM also tend to be flat.
Thus, throughout this study,
we work under the assumption that 
the magnonic Casimir energy of the AFM, NiO,
is dominated by
the two bands of Eq.~\eqref{eqn:dispersion}.
For a more accurate estimation,
inelastic neutron scattering measurement of 
its higher energy bands is essential.

\

\

{\section{Edge or surface magnon modes}
\label{sec:EdgeSurface}
}

We add an explanation about 
the effect of edge or surface magnon modes
on the magnonic Casimir energy.
The magnonic Casimir effect in our setup 
(see the thin film of Fig.~\ref{fig:system}) 
is induced by quantum fields for magnons 
of wavenumbers $k_z$ discretized by small $N_z$:
Its necessary condition is a $k_z$-dependent dispersion relation
through the discretization of $k_z$.
In this study, 
we consider thin films of $ N_z \ll N_x, N_y $.
Even if 
there are edge or surface magnon modes,
they are confined only on the $xy$ plane,
and their wavenumber in the $z$ direction is always zero,
i.e., $k_z=0$,
where its energy dispersion relation is independent of $k_z$.
Therefore,
such edge or surface modes cannot contribute to 
the magnonic Casimir effect.
In this sense, 
the magnonic Casimir effect in our setup
(see Fig.~\ref{fig:system})
is not affected by 
the presence or absence of edge or surface magnon modes.

\

\

{\section{The effect of edge conditions}
\label{sec:EdgeCondition}
}

We add a remark on the edge condition.
Details of the edge condition,
such as the presence or absence of disorder,
may affect the boundary condition
for the wave function of magnons,
but the magnonic Casimir effect is little influenced
as long as one does not assume an ultrathin film
such as $N_z = 1,2,3$.
Even if there is a change 
in the magnon band structure near the edge
due to some reasons,
such as changed spin anisotropies,
the existence of the magnonic Casimir effect
remains valid
as long as 
its necessary condition 
(see Sec.~\ref{sec:EdgeSurface} of this Supplemental Material)
is satisfied.



\end{document}